# Understanding the role of threading dislocations on 4H-SiC MOSFET breakdown under high temperature reverse bias stress

P. Fiorenza[1,a,*], M. Alessandrino[2], B. Carbone[2], C. Di Martino[2], A. Russo[2], M. Saggio[2], C. Venuto[2,] E. Zanetti[2], F. Giannazzo[1], F. Roccaforte[1]

[1]Consiglio Nazionale delle Ricerche – Istituto per la Microelettronica e Microsistemi (CNR-IMM), Strada VIII, n.5 Zona Industriale, 95121 Catania, Italy

[2]STMicroelectronics, Stradale Primosole 50, 95121 Catania, Italy

[a]patrick.fiorenza@imm.cnr.it

***Abstract***

The origin of dielectric breakdown was studied on 4H-SiC MOSFETs that failed after three months of high temperature reverse bias (HTRB) stress. A local inspection of the failed devices demonstrated the presence of a threading dislocation (TD) at the breakdown location. The nanoscale origin of the dielectric breakdown was highlighted with advanced high-spatial-resolution scanning probe microscopy (SPM) techniques. In particular, SPM revealed the conductive nature of the TD and a local increase of the minority carrier concentration close to the defect. Numerical simulations estimated a hole concentration 13 orders of magnitude larger than in the ideal 4H-SiC crystal. The hole injection in specific regions of the device explained the failure of the gate oxide under stress. In this way, the key role of the TD in the dielectric breakdown of 4H-SiC MOSFET was unambiguously demonstrated.

INTRODUCTION

Silicon carbide (4H-SiC) is a suitable material for high power and high temperature electronics [1]. In particular, metal oxide semiconductor field effect transistors (4H-SiC MOSFETs) are desired in automotive applications, as they guarantee a better efficiency in power conversion [2]. In this context, the 4H-SiC MOSFETs instability during long term interdiction under high temperature reverse bias stress (HTRB) is object of intensive investigations [3,4,5].

Since 4H-SiC MOSFETs are typically fabricated using $SiO_2$ as a gate insulator, one would expect a breakdown kinetics similar to the $SiO_2$/Si system. However, the breakdown kinetics of the $SiO_2$/4H-SiC system is more complex. In fact, while the behaviour of thin thermal gate oxides obeys to the percolation theory [6,7], for oxide thickness exceeding 10 nm, a deviation from the percolation theory is observed and explained by the presence of carbon-related defects in the oxide [8,9].

Premature failure for $SiO_2$/4H-SiC system, i.e., breakdown at zero time, has been often attributed the presence of crystalline killer defects (pipes, carrots, bars, etc) [10], and to the step-bunching on the 4H-SiC surface [11,12]. However, the role of other crystalline defects is still controversial. In particular, it is still unclear whether threading dislocations (threading screw dislocations (TSDs) and threading edge dislocations (TEDs)) behave as killer defects in 4H-SiC MOSFETs. Indeed, these crystalline defects have been blamed [13,14] and absolved [15,16] of causing premature device breakdown.

It should be pointed out that literature studies on the breakdown kinetics are typically based on the characterization of small MOS capacitors, which are unlikely to represent the behaviour of real power MOSFETs (i.e., having a larger area and a different geometry). In addition, due to the nanoscale size of the TDs and the large area of power MOSFETs, only local analyses based on advanced high-spatial-resolution techniques can enable to explain the origin of the breakdown in real devices.

This paper reports a clear correlation between threading dislocations (TDs) and dielectric breakdown (BD) on 4H-SiC planar MOSFETs, by the use of several advanced nanoscale electrical and structural characterization techniques. In fact, since the TDs appear on the 4H-SiC as narrow spot-like defect having a radius < 100 nm, only high-spatial-resolution measurements can provide useful insights on the physics involved in the device failure. Furthermore, numerical TCAD simulations explained the impact of the TD in terms of electric field and minority carrier distribution.

EXPERIMENTAL DETAILS

Several hundreds of 4H-SiC MOSFETs, designed to operate at 650V, were stressed in a HTRB test at 140 °C and $V_{DS}$=600 V for $10^7$ s (three months). The population of devices that survived to the stress test was a large fraction (98%) of all the tested devices. The remaining 2% of failed devices – showing a high slope in the Weibull distribution – was studied to understand the origin of the long time HTRB "intrinsic" failure mechanisms.

The HTRB test was performed using an Infinity 1000 Qualitau Equipment, able to collect simultaneously the current flows for each device. This capability is fundamental to distinguish the failed device and to record the information on each single BD event, defined from the abrupt increase of the gate current.

First, a PHEMOS1000 Emission microscopy (EMMI) was used to identify the coordinates of the breakdown event and a focussed ion dual beam (FIB) to mark the position of the breakdown. Afterward, the devices were completely delayered to expose the 4H-SiC bare surface. Nanoscale resolution current and capacitance mapping on the 4H–SiC surface were carried out by conductive atomic force microscopy (C-AFM) and scanning capacitance microscopy (SCM), respectively, using a DI3100 AFM by Veeco equipped with a Nanoscope V controller. Transmission electron microscopy (TEM) were performed using a Hitachi HD2300 STEM operated at 200 kV. Current - voltage (I–V) measurements were carried out using an Agilent B1505A parameter analyser. Numerical simulations of the electric field and carrier concentration were performed using Silvaco TCAD tools.

RESULTS AND DISCUSSION

Fig. 1a reports the gate current as a function of the HTRB stress time at 140 °C and $V_{DS}$=600 V, showing the abrupt increase of the current at the breakdown. Fig. 1b compares the $I_D$-$V_D$ characteristic of the failed device with that of a reference device survived to the HTRB test. As can be seen, the failed device shows a degradation of the output characteristics, as the drain current cannot be modulated by the gate bias after the HTRB. By repeating three times the $I_D$-$V_G$ ($V_{DS}$ = 0.5 V) transfer characteristics (Fig. 1c), the failed device shows a shift of the transfer characteristics, indicating the occurrence of a charge trapping. The presence of charge trapping is also responsible for the behaviour of the $I_G$-$V_G$ characteristic shown in Fig. 1d, where a significant increase of the gate current is observed at $V_G$>2V for the failed device. It is worth noting that during the HTRB some trapping phenomena occurred. In fact, the gate current at fixed bias (Fig. 1a) decreases with increasing the HTRB time, thus indicating the occurrence of holes trapping in the oxide layer until the hard breakdown occurs. In particular, during HTRB holes are accumulated at the $SiO_2$/4H-SiC interface and they can be injected in the oxide via Fowler-Nordheim tunnelling [17].

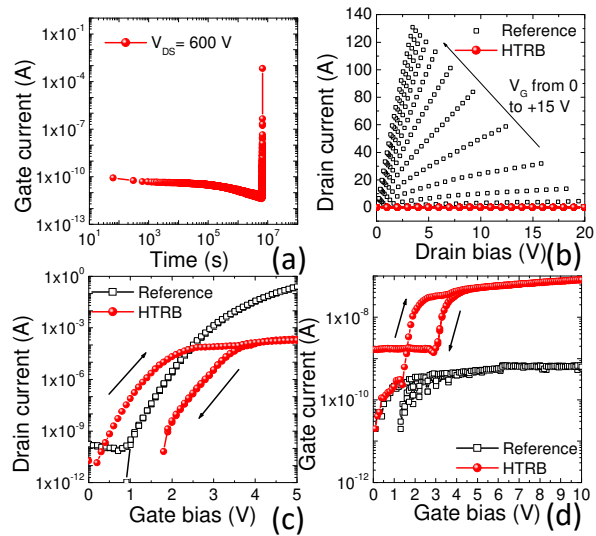

Fig. 1: (a) MOSFET gate current vs time under HTRB at $V_{DS}$ = 600 V at 140 °C. (b) Comparison of the $I_D$-$V_D$ collected on a fresh reference device and a failed HTRB device. (c) Comparison of hysteresis loop of the $I_D$-$V_G$ ($V_{DS}$ = 0.5 V) collected on a fresh reference device and a failed HTRB device. (d) Comparison of hysteresis loop of the $I_G$-$V_G$ collected on a fresh reference device and a failed HTRB device.

After the HTRB test, the breakdown spot was identified by EMMI (Fig. 2a) and marked with a FIB cut. After the complete device delayering, the bare 4H-SiC surface in the FIB mark was investigated by SEM, revealing the presence of a triangular features having the vertex in the centre of the JFET region (Fig. 2b).

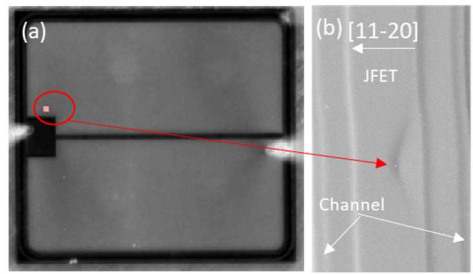

Fig. 2: (a) EMMI micrograph showing the location of the breakdown (BD) event. (b) SEM micrograph showing the delayered MOSFET, revealing the presence of epitaxial defect in the JFET region.

The nature of the defect found at the breakdown spot was clarified by means of TEM analysis. In fact, TEM in cross section (Fig. 3) allowed to identify a threading dislocation (TD) running along the epitaxial layer. The TEM dual beam investigation – along the directions [11-20] and [0002], shown in Figs. 3a and 3b, demonstrated the presence of a mixed dislocation. In fact, the dislocation shows up with the [11-20] spot (edge) and with the [0002] spot (screw).

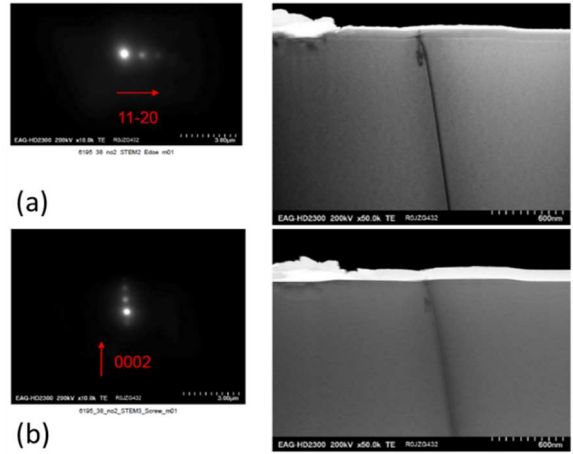

Fig. 3: (a) Cross section transmission electron microscopy in 2-beams configuration [11-20] (b) Cross section TEM in 2-beams configuration [0002]

To explore the electrical behaviour of the TD, a nanoscale electrical characterization was carried out using the conductive atomic force microscopy (C-AFM) [18]. In particular, C-AFM allowed determining the morphological shape of the surface (Fig. 4a) – a triangle with the vertex in the [11-20] direction – about 25 nm deeper than the surface of the JFET region where it is located. Even in the triangular region (highlighted with a dashed line), the surface conductivity is rather homogeneous (Fig. 4b). Noteworthy, the current is at least two orders of magnitude larger on the isosceles triangle vertex, i.e., where the TD is reaching the surface (Fig. 4b). Hence, the electronic transport properties close to the TD (radius < 100 nm) are significantly different from the surrounding material.

Then, local capacitance mapping [19] was performed using the SCM. During the surface scan with the metal tip, a modulating bias with amplitude ΔV at 100 kHz frequency is applied to the sample and the capacitance variation ΔC in response to this modulation is recorded with the SCM sensor. Besides the SCM signal amplitude |ΔC|, which is related to the net active dopants concentration ($N_A$-$N_D$) in the semiconductor underneath the tip, also the phase signal is recorded, which is very sensitive to the type of majority carriers in the region underneath the tip. Fig. 4c and d show the maps of the ΔC amplitude and phase measured in the same region where the C-AFM (Fig. 4b) was collected. The amplitude is not affected by the presence of the defect, thus indicating no significant variation of dopants concentration in the defect region. On the other hand, a difference in the in the phase map is observed between the triangular region and the rest of the n-type doped JFET area. As discussed in the following, this change in the phase signal can be ascribed to an increase of the density of minority carriers (holes) in this region.

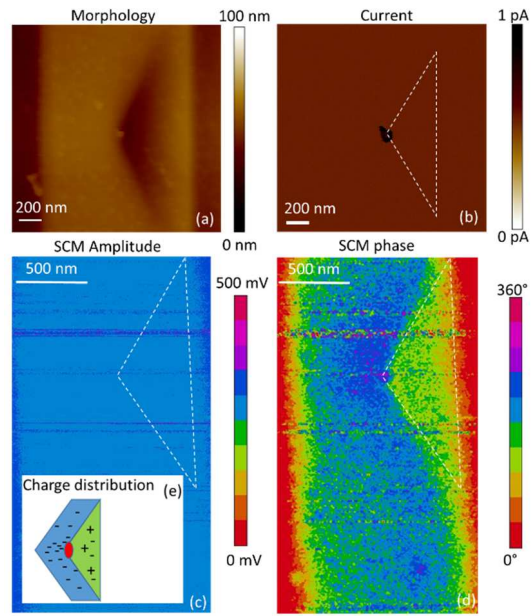

Fig. 4: (a) Atomic force microscopy morphology in the JFET region (b) C-AFM Current map showing the increased conduction in the TD. SCM amplitude (c) and phase (d) showing a spatial charge variation in the triangular defect region. Schematic of the local distribution of minority carriers (e).

Using the *Read's model* [20], *Chung et al* [21] considered the TD in the JFET n-type semiconductor region as a core containing acceptor-type states capable of trapping electrons, leading to a negatively charged dislocation core. They supposed the charge screened by a positive space charge in a cylindrical configuration of ionized donors. However, according with the SCM experimental data, the real scenario seems to be more complex. Fig. 4e schematically describes the non-cylindrical space charge distribution found in the TD region. On the left side – in the centre of the JFET region – an electron accumulation (blue colour) is demonstrated. On the other hand, inside the triangular region (green colour), a hole (minority carrier) accumulation produces a SCM phase variation that maintains the charge neutrality. Clearly, the TD produces a variation in the local electronic structure of the JFET region of the MOSFET.

The electronic structure of the TD has been theoretically investigated in literature [22,23]. *Łazewski et al* [23] employed density functional theory methods to calculate the energy levels and the effective band gap of the TD. Accordingly, inside the TD some additional localised broad states appeared, producing an effective reduction of the magnitude of the insulating gap. In particular, the TD possesses an identical conduction band edge but the effective band gap results approximately one forth smaller than in 4H-SiC, i.e., the valence band edge of the TD is 0.8 – 1 eV higher than in 4H-SiC. We used this information to quantify the impact of a TD located in the centre of the JFET of our MOSFET in HTRB configuration.

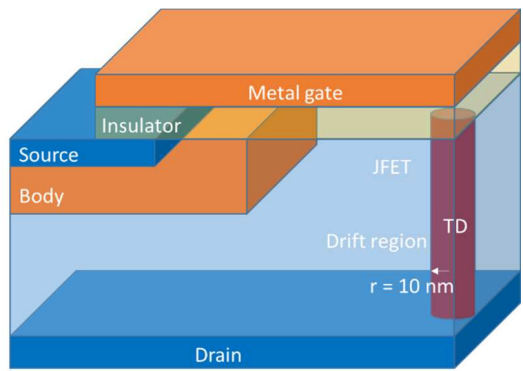

Fig. 5: Schematic 3D of the elementary cell of the MOSFET including the treading dislocation (TD) in the centre of the JFET region.

Fig. 5 shows the 3D schematic of a power MOSFET having a cylindrical TD of radius 10 nm in the centre of the JFET region. The band gap of the TD is assumed to be 2.3 eV with a common $E_C$ value but with a $E_V$ 1 eV higher than the defect-free 4H-SiC. Fig. 6a shows the cross-section on the electric field distribution inside the MOSFET unitary cell in HTRB configuration (at $V_{DS} = 600$ V), calculated using TCAD simulations. A cross section on the $SiO_2$ gate insulator at 25 nm from the physical interface (Fig. 6b) shows that the oxide electric field increases from the channel region toward the JFET region reaching its maximum (~ 4.6 MV/cm) at the edge of the unitary cell. To understand the TD behaviour inside the JFET region, the oxide electric field was calculated in both the ideal (reference) structure and in the structure containing the TD. A zoom in the centre of the JFET region reveals an increase of the oxide electric field of about 0.3% with respect to the reference ideal structure (inset in Fig. 6b). The higher valence band offset and the lateral confinement of the TD produces the creation of a quantum well. Considering the ideal conduction contribution – i.e., the ideal Fowler-Nordheim (FN) tunnelling [24] and the quantum confinement that inhibits the FN tunnelling from the bottom of the quantum well [25] – the oxide electric field produces a minor increase of the hole current through the insulator layer. Besides the calculated increase of the electric field due to the TD band gap narrowing, also the surface morphology plays a role [26]. However, this local increase of the oxide electric should not be prominent, as an acceleration of the breakdown kinetics would result in the MOSFET failure in a short time.

The band gap narrowing produces a quantum well for holes ("hot spot") that are collected from the surrounding 4H-SiC material (Fig. 6c). Hence, TD quantum well becomes a source for hot holes injection into the insulator (Fig. 6c). In fact, the calculated intrinsic carrier concentration ($n_i$) and holes concertation ($n_h$) increase in the TD of several orders of magnitude. Fig. 6d shows that the electron density ($n_e$) increases in the MOSFET profile following the electric field profile (Fig. 6b) while an abrupt increase of 13 and 9 orders of magnitude for $n_h$ and $n_i$ respectively is found. The local holes confinement can accelerate the oxide degradation. This demonstrates that the premature BD is not due to an intrinsic insulator weakness but the local variation of the semiconductor electronic properties in the presence of the TD in the JFET region.

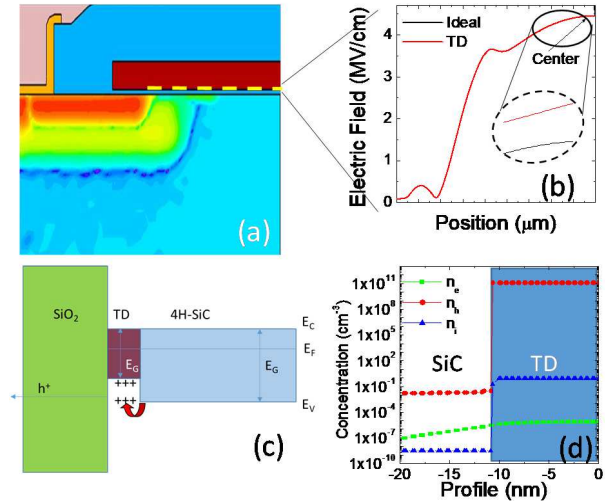

Fig. 6: (a) Cross section of the computed (SILVACO) MOSFET elementary cell. (b) Electric field profile in the middle of the gate insulator (dashed yellow line in Fig. 6a). (c) Schematic band diagram showing the holes accumulation in the TD and the injection in the SiO$_2$. (d) Carrier concentration profile across the TD and the ideal 4H-SiC.

CONCLUSION

Nanoscale structural and electrical investigations were carried out on large area power MOSFETs to clarify the role of the threading dislocation in the 4H-SiC epitaxial layer on the dielectric breakdown under high temperature reverse bias operation. The nanoscale inspection at the breakdown location of failed devices always revealed the presence of a threading dislocation. Hence, we can conclude that threading dislocations have a major role on the origin of the breakdown. Furthermore, C-AFM revealed an enhanced conductivity inside the TD. High-spatial-resolution SCM revealed a triangular spatial charge distribution surrounding the TD, compatible with a local band gap narrowing and an increase on the minority carrier concentration, 13 orders of magnitude larger than in the 4H-SiC bulk. Hence, the TD behaves as hot spot with a high concentration of holes that accelerate the breakdown kinetics.

ACKNOWLEDGMENTS

This work was carried out in the framework of the ECSEL JU project REACTION (first and euRopEAn siC eigTh Inches pilOt liNe), grant agreement no. 783158.

ORCIDs

Patrick Fiorenza https://orcid.org/0000-0002-9633-7892

Filippo Giannazzo https://orcid.org/0000-0002-0074-0469

Fabrizio Roccaforte https://orcid.org/0000-0001-8632-0870